\documentclass[twocolumn,aps,prb]{revtex4}
\usepackage{graphics,amsmath}
\usepackage{psfig}

\begin{document}

\title{Strong-coupling properties of unbalanced Eliashberg superconductors}

\author{E. Cappelluti$^{1,2}$ and G.A. Ummarino$^3$}

\affiliation{$^1$SMC Research Center, INFM-CNR c/o ISC-CNR,
v. dei Taurini 19, 00185 Roma, Italy}

\affiliation{$^2$Dipartimento di Fisica,
Universit\`a ``La Sapienza'', P.le  A. Moro 2, 00185 Roma, Italy}

\affiliation{$^3$Dipartimento di Fisica and CNISM, Politecnico di Torino,
Corso Duca degli Abruzzi 24, 10129 Torino, Italy}

\begin{abstract}
In this paper we investigate the thermodynamical properties
of ``unbalanced'' superconductors, namely, systems where the
electron-boson coupling $\lambda$ is different in the self-energy
and in the Cooper channels. This situation is encountered in a variety
of situation, as for instance in $d$-wave superconductors.
Quite interesting is the case where the pairing in the self-energy
is smaller than the one in the gap equation. In this case we predict
a finite critical value $\lambda_c$ where
the superconducting critical temperature $T_c$ diverges
but the zero temperature gap is still finite.
The specific heat, magnetic critical field and the penetration depth
are also evaluated.
\end{abstract}
\date{\today}
\maketitle

\section{Introduction}

The Eliashberg's theory of superconductivity represents an elegant
and powerful formalism to extend the BCS theory to
real materials.
Main achievements of the Eliashberg's theory are the generalization
of the BCS scheme to the strong coupling regime, where the
dimensionless electron-phonon coupling constant $\lambda$ can be of the order
or larger than unit, $\lambda \gtrsim 1$, and the inclusion
of the retarded nature of the electron-phonon interaction,
characterized by the phonon energy scale
$\omega_{\rm ph}$.\cite{eliashberg}
Within this framework it was possible to understand and predict
a number of characteristic features of
the strong coupling regime, as a $2\Delta/T_c$ ratio
larger than the BCS limit, the temperature dependence
of the magnetic critical field and of the specific heat, the appearance
of phonon features for $\omega > \Delta$ in the tunneling
and in the optical spectra.\cite{scalapino,carbotte_rmp}
Some results of the Eliashberg theory have become widely used
paradigmatic milestones, as for instance
the employment of McMillan-like formulas,
$T_c \approx \exp[-(1+\lambda)/\lambda]$, to estimate
the critical temperature and its dependence on
the microscopic interaction in generic superconductors.\cite{allen-mitrovic}
The strong-coupling limit $T_c$, $\Delta \gg \omega_0$
of the Eliashberg's theory has also been examined
in details, showing a drastic change of the superconducting properties,
with, for instance,
$T_c$, $\Delta \propto \sqrt{\lambda}\omega_0$.\cite{carbotte_rmp,allen-dynes,kresin,dolgov}

Different evolutions of the Eliashberg's theory have also been later
introduced in the course of years to adapt it to the particular cases
of specific materials.
Multiband effects,\cite{suhl,moskalenko,kresin1,golubov,nicol}
anisotropy and
non $s$-wave symmetries of the order
parameter,\cite{allen-mitrovic,millis,rieck,carbotte,zaza,ummarino}
effect of vertex corrections\cite{psg,gps,gpsprl}
have been for instance considered.
In all these cases one should consider in principle
the possibility that the electron-phonon coupling
(or any kind of other mediator) can be substantially different
in the self-energy and in the superconducting Cooper channels.
This is most evident in the case of $d$-wave pairing.
For example, if we
assume a factorized interaction,
$\alpha^2F({\bf k},{\bf k'};\omega)=
\alpha^2F(\omega)\psi_d({\bf k})\psi_d({\bf k'})$,
where $\alpha^2F({\bf k},{\bf k'};\omega)$ is the generic anisotropic
Eliashberg's function, and $\psi_d({\bf k})$ is the wave-function
for the $d$-wave symmetry, we would get no contribution
in the self-energy. This is expected for instance
in the case of a spin-mediated coupling where the exchange
energy $J({\bf k},{\bf k'})$ is factorizable as
$\approx J \psi_d({\bf k})\psi_d({\bf k'})$,\cite{miyake,scalapino2}
and the characteristic energy scale of the pairing, $\omega_{\rm sf}$,
is given by the spin-fluctuation spectrum.
Of course this is an extreme limit, and in real systems
there will be finite contributions in both the self-energy
and the Cooper channels, although in principle arising from
different electron-boson modes.
In any case, there is no reason to expect that the
electron-phonon coupling $\lambda_Z$ relevant for the $Z(\omega)$ 
renormalization wave-function to be the same as the one $\lambda_\Delta$
ruling the gap equation.

In this paper we investigate in details the consequence
of a different coupling in the $Z(\omega)$ wave-function
and in the gap equations.
We define this situation as ``unbalanced'' Eliashberg theory.
We focus here on thermodynamical quantities
which can be evaluated in the Matsubara space.
Spectral properties, involving
analytical continuation on the real axis,
will be investigated in a future publication.
We show that, contrary to the common feeling,
an unbalanced coupling in the Eliashberg's theory
has important and drastic differences with respect to the conventional
Eliashberg phenomenology. 
In particular we show that for $\lambda_Z < \lambda_\Delta$ the
superconducting critical temperature $T_c$ is strongly enhanced for finite
values of $\lambda \approx 1$, and in the infinite bandwidth limit
$T_c$ even diverges.
We also show that these new features
are strictly related to the retarded nature of any boson interaction,
accounting for the fact that this phenomenology was never discussed
in the case of the non-retarded BCS theory.

\section{Critical temperature $T_c$ vs. $\lambda$}

Let us start by consider the Eliashberg's equations for
the simple representative case of an Einstein boson spectrum.
To simplify the notations, we define
$\eta=\lambda_Z/\lambda_\Delta$ the ratio between
the electron-boson coupling in the $Z$ and in the gap equations,
and we simply denote $\lambda=\lambda_\Delta$.
In the Matsubara space we have
\begin{eqnarray}
Z_n&=&
1+ \frac{\eta\lambda\pi T}{\omega_n}
\sum_m 
\frac{\omega_0^2}{\omega_0^2+(\omega_n-\omega_m)^2}
\frac{\omega_m}{\sqrt{\omega_m^2+\Delta_m^2}},
\label{Z}
\\
Z_n \Delta_n&=&
\lambda\pi T
\sum_m 
\frac{\omega_0^2}{\omega_0^2+(\omega_n-\omega_m)^2}
\frac{\Delta_m}{\sqrt{\omega_m^2+\Delta_m^2}},
\label{Delta}
\end{eqnarray}
where $\omega_0$ is the energy scale of a generic bosonic mediator.
Eqs. (\ref{Z})-(\ref{Delta}) can be easily generalized in the case
of a $d$-wave symmetry for the gap order parameter
$\Delta_n \rightarrow \Delta_n \psi_d({\bf k})$ in the Cooper
pairing.
In the weak-intermediate regime, defined as
$T_c/\omega_0$, $\Delta/\omega_0 \ll 1$, a simple analytical solution
for $T_c$ and $\Delta$ is provided
by the square-well model.\cite{allen-mitrovic}
Along this line one recovers,
according the common wisdom,
a generalized McMillan-like formula
$T_c \approx \exp[-(1+\lambda_Z)/\lambda_\Delta]$, which predicts an
upper limit for $T_c$ in this case as well as in the perfectly balanced
$\eta=1$ case.
The validity of such result is however limited to
the weak-intermediate case where
$T_c/\omega_0$, $\Delta/\omega_0 \ll 1$.
In the balanced case, for instance,
a careful analysis shows that, in the strong coupling regime
$T_c/\omega_0$, $\Delta/\omega_0 \gg 1$ the critical
temperature as well the superconducting gap
do not saturate for $\lambda \rightarrow \infty$ but
they scale asymptotically as
$T_c$,$\Delta \propto \sqrt{\lambda}
\omega_0$.\cite{carbotte_rmp,allen-dynes,kresin,dolgov}

A first insight that things can be radically different for
an unbalanced Eliashberg's theory comes from a reexamination
of the strong coupling regime.
Plugging Eq. (\ref{Z}) in (\ref{Delta}) we obtain for $T_c$:
\begin{eqnarray}
\Delta_n&=&
\lambda \pi T_c
\sum_m 
\frac{\omega_0^2}{\omega_0^2+(\omega_n-\omega_m)^2}
\nonumber\\
&&
\times
\left[
\frac{\Delta_m-\eta(\omega_m/\omega_n)\Delta_n}{|\omega_m|}
\right].
\label{Delta-Z}
\end{eqnarray}
For $\eta = 1$
the term $n=m$ vanishes in Eq. (\ref{Delta-Z}), so that,
for $T_c \gg \omega_0$,
the first contribution in the boson propagator comes
from $\omega_0^2/[\omega_0^2+(\omega_n-\omega_m)^2]\approx
[\omega_0^2/4\pi^2 T_c^2 (n-m)^2]$.\cite{carbotte_rmp}
This is no more the case
for the unbalanced case where the leading contribution comes from
$\omega_0^2/[\omega_0^2+(\omega_n-\omega_m)^2]\approx \delta_{n,m}$.
Eq. (\ref{Delta-Z}) reads thus:
\begin{eqnarray}
1
=
\frac{\lambda(1-\eta)}{|2n+1|}.
\label{tc}
\end{eqnarray}
Note that the temperature $T$ does not appear anymore
in Eq. (\ref{tc}).
For $\eta<1$ Eq. (\ref{tc}) implies
that there is an upper limit $\lambda^{\rm max}\sim 1/(1-\eta)$
above which the system is always unstable {\em at any temperature}
with respect to the superconducting pairing.
A detailed analysis (see Appendix \ref{app-asympt}) shows:
\begin{eqnarray}
T_c^{\eta<1}
&=&
\frac{\omega_0}{2\pi}
\sqrt{\frac{\lambda\lambda_c}{\lambda_c-\lambda}},
\label{lambdac}
\end{eqnarray}
where $\lambda_c=1/(1-\eta)$.
On the other hand, for $\eta > 1$, Eq. (\ref{tc}) is never fulfilled
signalizing that the limit $T_c \gg \omega_0$ is unphysical
and $T_c$ must saturate for $\lambda \rightarrow \infty$ limit.

\begin{figure}[t]
\centerline{\psfig{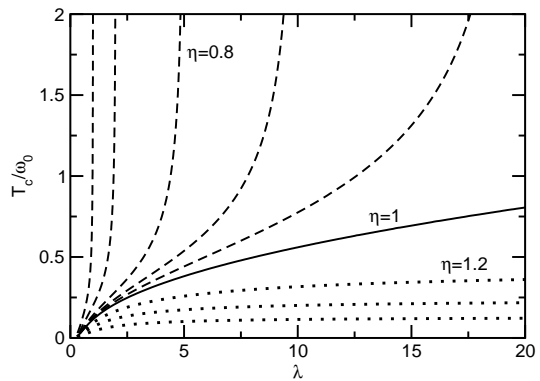}}
\caption{Critical temperature as function of
the pairing coupling $\lambda$ for different values of $\eta$.
Dashed lines: $\eta<1$; dotted lines: $\eta > 1$; the solid line
is the conventional Eliashberg result for $\eta=1$. From upper to lower
line: $\eta=0, 0.5, 0.8, 0.9, 0.95, 1, 1.2, 1.5, 2$.}
\label{f-tc}
\end{figure}

We would like to point out that the analytical divergence
$T_c \rightarrow \infty$ for $\lambda \rightarrow \lambda_c$
is strictly related to the infinite bandwidth model employed
in Eqs. (\ref{Z})-(\ref{Delta}).
On the other hand, in physical systems
the presence of a finite bandwidth $W$
determines an additional regime, $T_c \gg W$, where the analytical
divergence of $T_c$ at $\lambda_c$ is removed and
$T_c \propto \lambda W$ (Appendix \ref{app-asympt}).
In this respect the bandwidth $W$ defines an upper limiting regime for $T_c$.
Since in physical systems, however, $W$ is some orders of magnitude
bigger than the bosonic energy scale $\omega_0$, in the following,
for sake of simplicity, we shall concentrate on the infinite bandwidth limit
$W \gg T_c, \Delta, \omega_0$, keeping in mind however that
the analytical divergences found in this case will be removed when
finite bandwidth effects are included in the regime $T_c \gtrsim W$.

In Fig. \ref{f-tc} we show the critical temperature $T_c$
as function of the electron-boson interaction $\lambda$
obtained from the numerical solution
of Eqs. (\ref{Z})-(\ref{Delta}). We see that the conventional
Eliashberg case $\eta=1$, where $T_c$ scales as $\sqrt{\lambda}$,
represents rather an exception than the rule: for $\eta<1$
$T_c$ diverges at finite values of $\lambda$ determining, for each $\eta$,
a upper value of $\lambda$ above which the system
is superconducting at any temperature, while, for $\eta>1$, $T_c$
saturates for $\lambda \rightarrow \infty$ at some value which also is
dependent on $\eta$.
We can estimate in this case
(see Appendix \ref{app-asympt}) an upper limit for $T_c$:
\begin{eqnarray}
T_{c,\rm max}^{\eta>1}
&=&
\frac{\omega_0}{2\pi\sqrt{\eta-1}}.
\label{tcmax}
\end{eqnarray}

Before to proceed on, we would briefly comment
on the $T_c$ divergence at finite $\lambda$ for $\eta<1$.
This result seems to contradict apparently the BCS scenario
$T_c \propto \exp(-1/\lambda)$
which can be thought as the extremely unbalanced $\eta=0$ case
where the one-particle renormalization processes are disregarded.
However, a closer look at Eq. (\ref{Delta-Z}) shows
that a fundamental role in deriving Eqs. (\ref{tc})-(\ref{lambdac})
is played by the proper treatment of the retarded nature of
the electron-boson interaction, which gives rise to the correlation
between $\omega_n$ and $\omega_m$ within the energy window $\omega_0$.
In this sense, neglecting the $Z$-function for $\eta=0$
in Eqs. (\ref{Z})-(\ref{Delta}) corresponds to a ``retarded BCS'' theory.
This is quite different from the usual conventional BCS framework where
the interaction is supposed to be non-retarded and the frequencies
$\omega_n$ and $\omega_m$ are uncorrelated. This scenario can be
achieved in the retarded BCS context only in the limit
$\omega_0 \rightarrow \infty$, which enforces the
$T_c/\omega_0 \rightarrow 0$ limit, namely, the weak-coupling regime.

\section{Superconducting gap  $\Delta_{\rm M}$ vs. $\lambda$}

\begin{figure}[t]
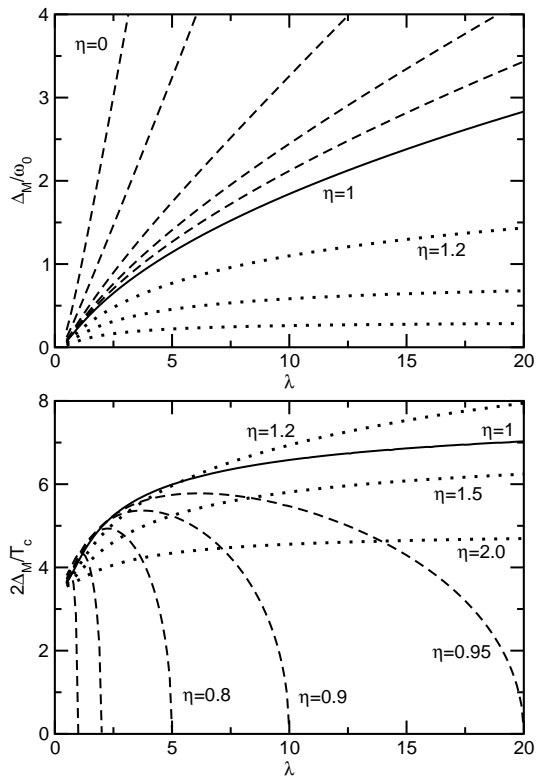

\centerline{\psfig{figure=f-gap.eps,width=7cm,clip=}}
\centerline{\psfig{figure=f-ratio.eps,width=7cm,clip=}}
\caption{Matsubara superconducting gap $\Delta_{\rm M}$ (upper panel)
and ratio $2\Delta_{\rm M}/T_c$ (lower panel)
as function of
the pairing coupling $\lambda$ for the same values of $\eta$
as in Fig. \ref{f-tc}}
\label{f-gap}
\end{figure}

Interesting enough, the $\Delta$ vs. $\lambda$ behavior
in an unbalanced superconductor can be quite different
from the $T_c$ vs. $\lambda$. In Fig. \ref{f-gap} we plot
the Matsubara superconducting gap $\Delta_{\rm M}$, defined as
$\Delta_{\rm M}=\lim_{T \rightarrow 0} \Delta_{n=0}$,
and the ratio $2\Delta_{\rm M}/T_c$
as function of the electron-boson coupling $\lambda$.
We remind that, although $\Delta_{\rm M}$ underestimates
the physical gap edge 
obtained by the analytical continuation on the real axis, 
the analytical dependence of these two quantities is usually the same,
so that $\Delta_{\rm M}$ can be reasonable employed to study
the limiting behavior of the superconducting gap
in the strong-coupling regime.
Detailed investigations on the real axis are however
needed to assess in a more formal way this issue.
Fig. \ref{f-gap} shows that,
while for $\eta > 1$ $\Delta_{\rm M}$ has a saturating behavior
similar as $T_c$, in the case $\eta < 1$
the superconducting gap does not diverge at some finite value
of $\lambda$, as $T_c$,
but rather increases linearly with the electron-boson coupling.
This different behavior can be also understood by applying
some analytical derivations properly generalized for
unbalanced superconductors.\cite{cmm}
Assuming $\Delta_{\rm M}\gg \omega_0$, and following Eqs. (4.29)-(4.35)
of Ref. \onlinecite{carbotte_rmp}, the superconducting gap
$\Delta_{\rm M}$ is determined by the following relation:
\begin{eqnarray}
1+\eta \lambda\frac{\pi \omega_0}{2 \Delta_{\rm M}}
-c_1\eta \lambda\frac{\omega_0^2}{\Delta_{\rm M}^2}
&=&
\lambda\frac{\pi \omega_0}{2 \Delta_{\rm M}}
-c_2\lambda\frac{\omega_0^2}{\Delta_{\rm M}^2},
\label{gapoo}
\end{eqnarray}
where $c_1$, $c_2$ are constant factors 
whose value is discussed in Appendix \ref{app-asympt}.
For $\eta=1$ the terms $\propto \omega_0/\Delta_{\rm M}$
on both the left and right sides cancel out,
so that $\Delta_{\rm M} \propto
\sqrt{\lambda} \omega_0$.\cite{cmm,carbotte_rmp}
This is no longer true for $\eta \neq 1$.
In particular, for $\eta < 1$ we find
\begin{eqnarray}
\Delta_{\rm M}
&=&
\lambda(1-\eta)\frac{\pi \omega_0}{2},
\end{eqnarray}
while, for $\eta> 1$, Eq. (\ref{gapoo}) does not admit
solution signalizing, once more, that the initial
assumption $\Delta_{\rm M} \gg \omega_0$ is inconsistent
in this limit and that $\Delta_{\rm M}$ must be saturate
for $\lambda \rightarrow \infty$.
By taking into account higher order terms
we found an upper limit $\Delta_{\rm M}$ for $\eta > 1$
in the regime $\lambda \rightarrow \infty$ in similar
way as done for $T_c$:
\begin{eqnarray}
\Delta_{\rm M, max}
&=&
\frac{4(\eta c_1-c_2)\omega_0}{2\pi(\eta-1)}.
\label{deltamax}
\end{eqnarray}
Note that $\Delta_{\rm M, max}$ in Eq. (\ref{deltamax})
diverges as $1/(\eta-1)$ whereas $T_{c,{\rm max}}$
in Eq. (\ref{tcmax}) scales as $T_{c,{\rm max}} \sim 1/\sqrt{\eta-1}$.
This means that the ratio $2\Delta_{\rm M}/T_c$ is not bounded
for $\eta >1$ and it can be even larger than
in the Eliashberg case $\eta=1$ and formally diverging
for $\eta \rightarrow 1$, in agreement with the numerical results
shown in Fig. \ref{f-gap}.

\section{Temperature dependence of $\Delta_{\rm M}(T)$}

In the previous sections we have studied the strong coupling behaviors
of the critical temperature $T_c$ and of the zero
temperature Matsubara gap $\Delta_{\rm M}$
in the limit $\lambda \gg 1$.
We have seen for instance that, in the $\eta < 1$ case,
$T_c$ diverges at some critical value $\lambda_c$, so that
for $\lambda > \lambda_c$ the system
is superconducting at any temperature.
This behavior is in contrast with the one of the Matsubara gap
$\Delta_{\rm M}$ which is always finite for any $\lambda$ and scales
linearly with $\lambda$ for $\eta < 1$.
As shown in Fig. \ref{f-gap}, these different behaviors are reflected
in a ratio $2\Delta_{\rm M}/T_c$ smaller than the BCS limit 3.53 and vanishing
for $\lambda \rightarrow \lambda_c$.
In this situation an interesting issue to investigate is the temperature
dependence of the superconducting gap $\Delta(T)$, which is reflected
in a number of observable physical behaviors, as the temperature
profile of the magnetic field $H_c(T)$, of the
London penetration depth $\lambda_{\rm L}(T)$
or of the specific heat $C_V(T)$.
Also intriguing is the situation $\eta < 1$ and $\lambda > \lambda_c$, where
a finite superconducting gap exists at zero temperature but where
no finite critical temperature is predicted. In this case
the temperature behavior of the gap itself is not clear and needs to be
investigated.

\begin{figure}[t]
\centerline{\psfig{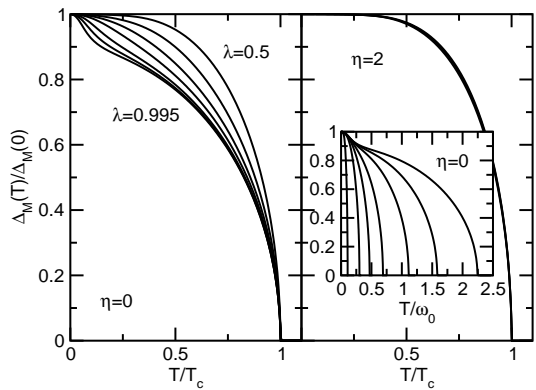}}
\caption{Temperature evolution of the Matsubara superconducting gap 
$\Delta_{\rm M}(T)/\Delta_{\rm M}(0)$
for different unbalanced cases. Left panel:
$\eta=0$ ($\lambda_c=1$)
and different coupling (from top to bottom)
$\lambda=0.5, 0.8, 0.9, 0.95, 0.98, 0.99, 0.995$.
Right panel:
$\eta=2$ and (from bottom to top, but barely
distinguishable) $\lambda=5, 10, 20$.
Inset: same quantities as in the left panel ($\eta=0$) but
as function of $T/\omega_0$.}
\label{f-gap_T}
\end{figure}

In Fig. \ref{f-gap_T} we show the temperature dependence of
$\Delta_{\rm M}(T)/\Delta_{\rm M}(0)$
(defined as $\Delta_{\rm M}(T)=\Delta_{n=0}$)
for different characteristic cases, namely
for $\eta=0$ and different $\lambda < \lambda_c$, and for $\eta =2$
and different $\lambda \gtrsim 1$. 
Most regular is the $\Delta_{\rm M}(T)$ vs. $T$ dependence for $\eta>1$,
where $\Delta_{\rm M}(T)$ follows a conventional
behavior, independently of the coupling $\lambda$. This
regular behavior can be understood by reminding that for $\eta > 1$,
even for very large coupling $\lambda$, the values
of $T_c$ and of the superconducting gap $\Delta_{\rm M}(T)$ are always finite
and (at most) of the same order of the energy $\omega_0$.
Quite different is the case of $\eta<1$,
here represented by $\eta=0$, where $\Delta_{\rm M}(T)$
shows a temperature dependence remarkably different from the
BCS one. For $\lambda$ close to $\lambda_c$,
in particular,
the superconducting Matsubara gap has a first initial
drop followed by a more regular dependence.
This change of curvature represents the crossover between a small temperature
($T/\omega_0 \lesssim 1/4$) to a large temperature
($T/\omega_0 \gtrsim 1/4$) regime, as shown in the inset of
Fig. \ref{f-gap_T} where we plot $\Delta_{\rm M}(T)/\Delta_{\rm M}(0)$
as function of $T/\omega_0$.
Note that, while the value of the critical temperature $T_c$ is
strongly dependent on the coupling $\lambda$, the initial dependence
of $\Delta_{\rm M}(T)$ is only weakly dependent on $\lambda$.
We remind indeed that the $T_c$ divergence for $\lambda \rightarrow \lambda_c$
is essentially a by-product of having reached the $T/\omega_0\gg 1$
for a  finite $\lambda$ in the unbalanced case.
We can thus understand the results of Fig. \ref{f-gap_T}
in the following way: for low temperature ($T/\omega_0 \lesssim 1/4$)
the superconducting gap probes a pairing kernel which is actually
increased by the lack of unbalance, but still regular,
(remind that $\Delta_{\rm M}(T=0)$, contrary to $T_c$, does
not diverge at $\lambda_c$, but it steadily increases as $\propto \lambda$
in the strong-coupling regime).
For low temperature $\Delta_{\rm M}(T)/\Delta_{\rm M}(0)$ follows
thus a standard-like behavior which would close to some finite $T'_c$
{\em not diverging} at $\lambda_c$.
When $\lambda$ is however close enough to $\lambda_c$, the range
$T/\omega_0 \gtrsim 1/4$ is achieved before $T'_c$ is actually reached;
in this regime high temperature effects become dominant
in the pairing kernel, reflected in a change of the
$\Delta_{\rm M}(T)$ vs. $T$ trend and in a final, physical, 
$T_c$ which diverges at $\lambda \rightarrow \lambda_c$.

It is interesting to investigate also how the superconducting
gap  $\Delta_{\rm M}(T)$ closes at $T_c$. 
In the conventional, perfectly balanced, Eliashberg theory
the normalized gap $\Delta(T)/\pi T_c$
(or equivalently $\Delta(T)/\Delta(T=0)$) scales indeed for
$T \rightarrow T_c$ as
$(\Delta(T)/\pi T_c)^2 \sim c \delta$,
where $\delta=1-T/T_c$ and where $c$ is a finite constant
which, in the weak-coupling BCS limit
$\lambda \ll 1$, is $c=0.95$ whereas for
$\lambda \gg 1$ one gets $c=2$.
This scenario is qualitatively different in the case
of unbalanced $\eta < 1$ superconductors
(Fig. \ref{f-gap_T}, left panel)
where the constant $c$
strongly depends on the coupling $\lambda$.

For $\eta < 1$ a first insight about the temperature dependence
of $\Delta_{\rm M}(T)$ close at $T_c$ is gained simply by
considering that, for $\lambda \rightarrow \lambda_c$,
$\Delta_{\rm M}(0)$ is finite while $T_c\rightarrow \infty$,
with a ratio $2\Delta_{\rm M}(0)/T_c \rightarrow 0$.
This means that, as $\lambda \rightarrow \lambda_c$,
the constant $c$ must vanish.
This result can be shown in a more quantitative way
(see Appendix \ref{app-asympt})
by employing a one-Matsubara-gap approximation\cite{carbotte_rmp}
which,
for $T \simeq T_c$ and for $T_c \gg \omega_0$, is
quite reasonably justified.
For generic $\eta$ one obtains thus
\begin{eqnarray}
\frac{\Delta_0^2}{\pi^2T_c^2}
&=&
4\delta
\frac{1-\lambda(1-\eta)}{2-\lambda(1-\eta)},
\label{gap-t}
\end{eqnarray}
where $\delta=1-T/T_c$.
Note that Eq. (\ref{gap-t})
reduces to the standard relation
$(\Delta_0/\pi T_c)^2=2\delta$ for $\eta=1$.\cite{carbotte_rmp}
On the other hand, for $\eta<1$
\begin{eqnarray}
\frac{\Delta_0^2}{\pi^2T_c^2}
&=&
4\delta
\frac{\lambda_c-\lambda}{2\lambda_c-\lambda},
\label{gap-t-1}
\end{eqnarray}
showing that the coefficient $c$ vanishes as
$c \propto \lambda_c-\lambda$ 
for $\lambda\rightarrow \lambda_c$.

\section{Other thermodynamical quantities}

The anomalous temperature dependence of $\Delta_{\rm M}(T)$ is reflected
also in other thermodynamical, measurable quantities, as the specific heat
$C_V(T)$ or the magnetic critical field $H_c(T)$.
In order to investigate these properties we evaluate numerically
the free energy difference $\Delta F$ between the superconducting
and the normal state,\cite{carbotte_rmp}
\begin{eqnarray}
\Delta F(T)
&=&
-\pi T N(0)\sum_n
\left(
Z^{\rm S}_n-\frac{Z^{\rm N}_n|\omega_n|}{\sqrt{\omega_n^2+\Delta_n^2}}
\right)
\nonumber\\
&&\times
\left(\sqrt{\omega_n^2+\Delta_n^2}-|\omega_n|\right),
\label{free}
\end{eqnarray}
where
$N(0)$ is the electron density of states at the Fermi level and
$Z^{\rm S}_n$, $Z^{\rm N}_n$ are the $Z$-renormalization functions
calculated respectively in the superconducting and in the normal state.
From Eq. (\ref{free}) we obtain the 
magnetic critical field $H_c(T)=\sqrt{-8\pi \Delta F(T)}$ and
the specific heat difference as
$\Delta C_V(T) = -T\partial^2 \Delta F/\partial T^2$.
One can also obtain the total specific heat
$C_V(T)=\Delta C_V(T)+C_V^{\rm N}$ by adding
the contribution that the system would have in the normal state,
$C_V^{\rm N}=\gamma T$, where $\gamma=(2/3)\pi^2N(0)(1+\eta \lambda)$
is the Sommerfeld constant.
In similar way we can evaluate
the London penetration depth $\lambda_{\rm L}(T)$ as:\cite{carbotte_rmp}
\begin{eqnarray}
\frac{1}{e^2 v_{\rm F}^2N(0)\lambda_{\rm L}^2(T)}
&=&
\frac{2\pi T}{3}
\sum_n
\frac{\Delta_n^2}{Z_n\left[\omega_n^2+\Delta_n^2\right]^{3/2}},
\label{london}
\end{eqnarray}
where $e$ is the electron charge and $v_{\rm F}$ the Fermi velocity.

\begin{figure}[t]
\centerline{\psfig{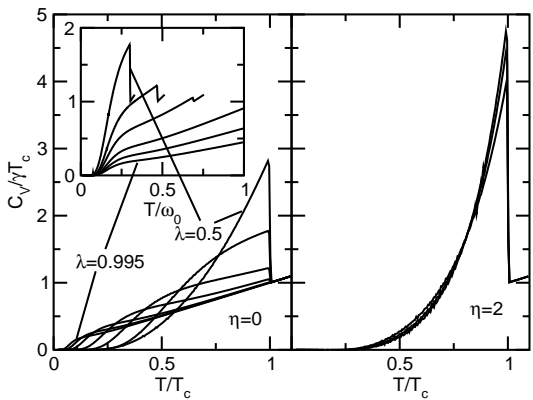}}
\caption{
Specific heat $C_V(T)$
for the two characteristic unbalanced cases
previous considered.
Left panel:
$\eta=0$ ($\lambda_c=1$)
and $\lambda=0.5, 0.8, 0.9, 0.95, 0.98, 0.99, 0.995$.
Right panel:
$\eta=2$ and $\lambda=5, 10, 20$.
Inset: specific heat for $\eta=0$
as function of $T/\omega_0$.}
\label{f-cv}
\end{figure}

In Fig. \ref{f-cv} we show the temperature dependence of the specific heat
as function of $\lambda$
for the two representative cases $\eta=0$, $\eta=2$.
In this latter case the specific heat has a regular activated behavior
and it is almost independent of $\lambda$,
in agreement with the corresponding weakly $\lambda$ dependence
of the temperature behavior of
$\Delta_{\rm M}(T)$ reported in Fig. \ref{f-gap_T}.
Note also that
for this value of $\eta$ the asymptotic value $\lambda \rightarrow \infty$
of the specific heat jump
$\Delta C_V/\gamma T_c \simeq 4.2$, is quite larger than the BCS limit,
$(\Delta C_V/\gamma T_c)_{\rm BCS} \simeq 1.43$, pointing out
that the superconductors is in an effective strong-coupling limit.
We should remark
that the asymptotic value
$\lim_{\lambda \rightarrow \infty}\Delta C_V/\gamma T_c$
is actually dependent on the specific value of the
parameter $\eta > 1$.

Quite anomalous is also the temperature behavior for $\eta<1$.
In this case we see that approaching $\lambda \rightarrow \lambda_c$
the jump is remarkably reduced.
Eqs. (\ref{gap-t})-(\ref{gap-t-1}) can be used to estimate
the jump $\Delta C_V$ at $T_c$ of the specific heat.
Using once more the one-Matsubara gap approximation,
and employing a standard analysis,\cite{carbotte_rmp}
one can show that
the formal expression for the free energy difference
$\Delta F$ close to $T_c$ is just equal as in the standard case,
\begin{eqnarray}
\Delta F = 
-
\frac{N(0)(\pi T_c)^2}{2}
\left(1+\eta\lambda\frac{\omega_0^2}{4\pi^2T^2}\right)
\left(\frac{\Delta_0(T)}{\pi T_c}\right)^4.
\label{standard}
\end{eqnarray}
Plugging Eq. (\ref{gap-t}) and (\ref{lambdac}) in (\ref{standard}), and
using the standard derivation of $\Delta C_V$, we obtain thus
\begin{eqnarray}
\frac{\Delta C_V}{\gamma T_c}
&=&
\frac{24}{1+\eta\lambda}
\left(1+\eta\frac{\lambda_c-\lambda}{\lambda_c}\right)
\left(\frac{\lambda_c-\lambda}{2\lambda_c-\lambda}\right)^2,
\label{cv-1}
\end{eqnarray}
showing that, for $\eta < 1$, also 
$\Delta C_V/\gamma T_c$ as $(\Delta_0/\pi T_c)^2$
scales as $\propto \lambda_c-\lambda$ 
for $\lambda\rightarrow \lambda_c$.
Once more, Eq. (\ref{cv-1}) reduces to the standard analytical
result $\Delta C_V/\gamma T_c=12/\lambda$ for $\eta=1$ and
$\lambda \rightarrow \infty$.\cite{mwc}
It is interesting to note that the vanishing of
$\Delta C_V/\gamma T_c$ for 
$\lambda\rightarrow \lambda_c$ is mainly due to
the vanishing of $\Delta C_V \propto (\lambda_c-\lambda)^{3/2}$,
whereas $T_c \propto 1/\sqrt{\lambda_c-\lambda}$.
This means that, contrary to the perfectly balanced case $\eta=1$
where the vanishing of $\Delta C_V/\gamma T_c$ is driven by
$\gamma T_c \rightarrow \infty$ which overcomes
the divergence of
$\Delta C_V$, in the unbalanced $\eta<1$ case the
specific heat jump $\Delta C_V$ is itself vanishing.
Such observation points out
that, although $T_c$ is much higher,
the superconducting properties in the $T/\omega_0$ regime
of $\eta<1$ unbalanced superconductors are much weaker than the usual.

This scenario is once more outlined in Fig. \ref{f-cv},
where we see that  the vanishing of $\Delta C_V$ is accompanied
by the developing of a shoulder at $T \approx \omega_0/4$ (see inset).
Above this temperature the specific heat scales roughly
linearly with $T$ as a normal ungapped metal with a smaller and
smaller jump at $T_c$ as $\lambda \rightarrow \lambda_c$.

\begin{figure}[t]
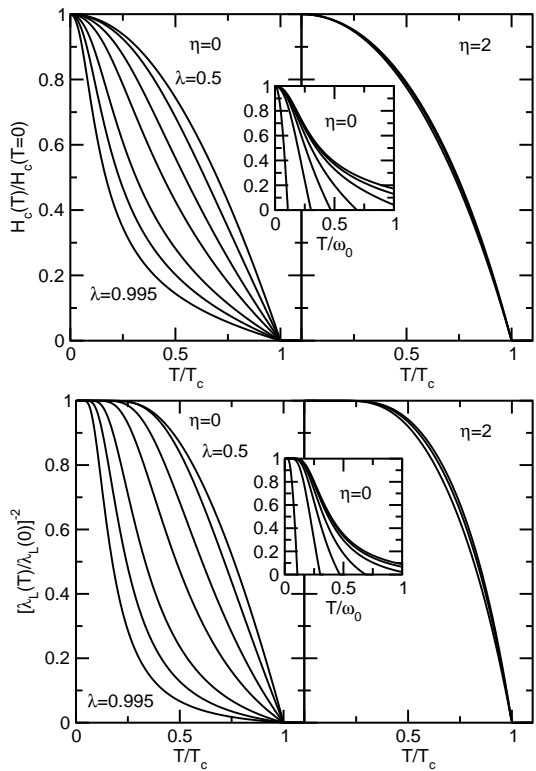

\centerline{\psfig{figure=f-hc_T.eps,width=7cm,clip=}}
\centerline{\psfig{figure=f-pene_T.eps,width=7cm,clip=}}
\caption{Temperature dependence of the
magnetic critical field $H_c(T)/H_c(0)$ and of the
London penetration depth $\lambda_{\rm L}^2(0)/\lambda_{\rm L}^2(T)$
for the two characteristic unbalanced cases
previous considered.
Left panels:
$\eta=0$ ($\lambda_c=1$)
and $\lambda=0.5, 0.8, 0.9, 0.95, 0.98, 0.99, 0.995$.
Right panels:
$\eta=2$ and $\lambda=5, 10, 20$.
Insets: the same quantities
as function of $T/\omega_0$.}
\label{f-hc}
\end{figure}

A similar trend is observed in the study of the
temperature dependence of the magnetic critical field $H_c$
and of the London penetration depth $\lambda_{\rm L}$,
shown in Fig. \ref{f-hc}.
For $\eta=2$ we find again that the temperature dependence
of both $H_c(T)$ and $\lambda_{\rm L}^{-2}(T)$ presents again
a conventional behavior with even a more marked curvature
with respect to the BCS curve. This is compatible with
the specific heat jump which is also larger than the BCS limit.
Quite different is the case for $\eta=0$. Here we observe a change
of curvature at low temperature which is more marked
as $\lambda \rightarrow \lambda_c$.\cite{note-hc}
Once again such change of curvature occurs for $T \gtrsim \omega_0/4$
and for $\lambda \simeq \lambda_c$ it is reflected in a sudden drop
of $H_c(T)$ and $\lambda_{\rm L}^{-2}(T)$ signalling once more
that, although the critical temperature is strongly increasing,
the high temperature
superconducting properties of unbalanced systems are quite poor.

\section{Summary and discussion}

In this paper we have investigated the properties of
``unbalanced'' retarded superconductors, namely systems where
the electron-boson coupling $\lambda_Z$
in the one-particle renormalization function
is different than the one
relevant for the gap equation, $\lambda_\Delta$.
We have shown that the superconducting properties are strongly
dependent on the ratio $\eta=\lambda_Z/\lambda_\Delta$.
We have analyzed both the cases $\eta>1$ and
$\eta<1$.
In the first case we show that, contrary to the perfectly balanced
case, the critical temperature is always finite
and it even saturates at a finite value for
$\lambda \rightarrow \infty$.
The superconducting properties in this case are quite similar
to the conventional case in the weak-intermediate regime
where the magnitude of $\eta$ rules the relevance of the
strong-coupling effects.
Quite anomalous is on the other hand the case $\eta<1$,
which is relevant in a variety of situations as for instance
for $d$-wave superconductors.
In this case
we find that the critical temperature,
in the infinite bandwidth limit, {\em diverges} at a finite
value $\lambda_c \approx 1$ ($\lambda_c=1$ for $\eta=0$).
We show also that, although for $\lambda>\lambda_c$ the critical
temperature $T_c$ is strongly enhanced for
$\lambda \rightarrow \lambda_c$, the zero temperature gap
is still finite and it scales as $\Delta \propto \lambda$.
The anomalous temperature dependence of the superconducting gap
is reflected in a variety of other physical properties, as
the magnetic critical field, the penetration depth and
the specific heat. All these quantities show a strong anomaly
at $T \gtrsim \omega_0/4$ above which the system presents very weak
superconducting properties, which however persist
up to the higher critical temperature $T_c$.
We would like to remark once more that
in real systems finite bandwidth effects remove the analytical divergence
of $T_c$ at $\lambda_c$ when $T_c \gg W$.

These results suggest an interesting pseudogap scenario.
In the $T_c' \lesssim T \lesssim T_c$ regime indeed,
since the superconducting binding energies are quite small,
the robustness of the long-range order of this
phase towards phase fluctuations, disorder,
phase separation and other different kinds of instability
is highly questionable. In case of loss of the long-range order, these
weak superconducting properties will present themselves as a pseudogap phase
in the range temperature $T_c' \lesssim T \lesssim T_c$, where the
temperature $T_c'$ will appear as the thermodynamical critical
temperature where true lost range order is lost, and
$T_c \approx T_c$ will set the pseudogap temperature $T^*$.
Note that, even in this case, the critical temperature 
of the long-range order will result $T_c' \approx \omega_0/4$,
which can be significantly higher than the predictions of the standard
Eliashberg's theory.

Before concluding, we would like to spend some more words
about the physical meaning
of the $T_c$ divergence for $\eta<1$.
On the mathematical ground, we have seen that, contrary to the conventional
balanced case, in unbalanced superconductors the Cooper instability
is driven by the $n=m$ term in Eq. (\ref{Delta-Z}).
In physical terms this corresponds to consider the {\em classical} limit
$T \gg \omega_0$ of the bosons.
This is quite different from the usual Eliashberg's theory where
only virtual bosons, which a characteristic energy scale $\omega_0$,
are responsible of the pairing. Just as for the
linear behavior $\rho(T) \propto \lambda T$ of the resistivity,
in the classical limit $T \gg \omega_0$
the energy scale $\omega_0$ does not provide anymore
any upper limit and the effective pairing is mainly ruled
by the increasing bosonic population $n(T) \propto T$
with temperature.
In the absence of any other competing effect,
increasing temperature will result thus in a stronger pairing
with a positive feedback which would lead
a high critical temperature $T_c \gg \omega_0$, and the only
limiting energy scale in this case is provided by the electronic bandwidth.
In this scenario
a competing effect in balanced superconductor is provided by the fact
that the increase of the boson population, as $T$ increases, would
act in a similar way in the self-energy channel. The competition
between these two effects give rise to the well-known
$T_c \sqrt{\lambda} \omega_0$ dependence in perfectly balanced
superconductors with $\eta=1$
Such equilibrium does not occur however
in unbalanced superconductors with $\eta<1$
where the gain in the Cooper channel is larger than the loss
in the self-energy. In this case, at sufficiently large $T$, the increase
of the Cooper pairing due to the boson population will prevail over
the one-particle renormalization effects and a superconducting ordering
can be sustained up to high temperatures $T_c$ limited only
by the electronic bandwidth energy scale.

\acknowledgments

It is a pleasure to thank C. Grimaldi, F. Marsiglio,
L.Pietronero, L. Benfatto, C. Castellani and M. Grilli
for many useful discussions.

\appendix
\section{Analytical formulas for asymptotic behaviors
$T_c \gg \omega_0$, $\Delta_{\rm M} \gg \omega_0$}
\label{app-asympt}

In this appendix we derive some useful limit
expressions for the critical temperature $T_c$ and
the Matsubara superconducting gap $\Delta_{\rm M}$
in the regimes
$T_c \gg \omega_0$, $\Delta_{\rm M} \gg \omega_0$.

\subsection{Critical temperature $T_c$}

Let us start by considering Eq. (\ref{Delta-Z}) and let us assume
$T_c \gg \omega_0$.
In this regime we can employ the one-Matsubara-gap
approximation,\cite{carbotte_rmp} where
the only not vanishing terms
of the Matsubara gap function $\Delta_n$
are $\Delta_{n=0}=\Delta_{n=-1}$.
We have thus:
\begin{eqnarray}
\Delta_0
&=&
\lambda\Delta_0+
\lambda\Delta_{-1}\frac{\omega_0^2}{4\pi^2T_c^2}
\nonumber\\
&&
-\lambda\eta \Delta_0
\sum_m
\frac{\omega_0^2}{\omega_0^2+(\omega_n-\omega_m)^2}
 \mbox{sgn}[\omega_m],
\label{ap1}
\end{eqnarray}
and, reminding that $\Delta_{-1}=\Delta_0$ and
$\sum_m \mbox{sgn}[\omega_m]
\omega_0^2/[\omega_0^2-(\omega_{n=0}-\omega_m)]=1$, we have
\begin{equation}
1=\lambda(1-\eta)+\lambda\frac{\omega_0^2}{4\pi^2T_c^2}.
\label{app-tc}
\end{equation}
For $\eta=1$ the first term on the right hand side
is zero, and we recover the usual result
$T_c^{\eta=1}=\sqrt{\lambda}\omega_0/(2\pi)$.
On the other hand, for $\eta<1$ we obtain
\begin{eqnarray}
T_c^{\eta<1}
&=&
\frac{\omega_0}{2\pi}\sqrt{\frac{\lambda}{1-\lambda(1-\eta)}},
\label{tc1}
\end{eqnarray}
or, equivalently,
\begin{eqnarray}
T_c^{\eta<1}
&=&
\frac{\omega_0}{2\pi}
\sqrt{\frac{\lambda_c\lambda}{\lambda_c-\lambda}},
\label{my}
\end{eqnarray}
where $\lambda_c=1/(1-\eta)$.

Finally, The same expression (\ref{tc1}) can be used to evaluated
an upper limit for $T_c$ in the case $\eta>1$.
We can write thus
\begin{eqnarray}
T_c^{\eta>1}
&=&
\frac{\omega_0}{2\pi}\sqrt{\frac{\lambda}{1+\lambda(\eta-1)}},
\end{eqnarray}
and, for $\lambda \rightarrow \infty$,
\begin{eqnarray}
T_{c,{\rm max}}^{\eta>1}
&=&
\frac{\omega_0}{2\pi\sqrt{\eta-1}}.
\label{tcmax_etabig}
\end{eqnarray}
Note that Eq. (\ref{tcmax_etabig}) predicts a saturating upper
limit for $T_c^{\eta>1}$ for $\lambda \rightarrow \infty$,
violating the assumption $T_c \gg \omega_0$.
This value, $T_{c,{\rm max}}^{\eta>1}$, can be thus assumed
as a upper limit for the asymptotic behavior of
$T_c^{\eta>1}$ in the regime $\lambda \rightarrow \infty$.

Let us now consider finite bandwidth effects in the $\eta<1$ case.
For simplicity we shall focus on the representative case $\eta=0$.
Finite bandwidth effects can be included
in the linearized Eliasberg's equations, (\ref{Delta-Z}), by replacing
the term $\pi/|\omega_m| \rightarrow (2/\omega_m)\arctan(W/2\omega_m)$,
where $W/2$ is the half-bandwidth in a symmetric particle-hole system.
In this case, in order to obtain an analytical expression for $T_c$
in the limit $T_c \gg \omega_0$, it is sufficient to retain
only the $n=m$ term in Eq. (\ref{Delta-Z}), 
and, in the one-Matsubara-gap approximation, we obtain
\begin{eqnarray}
\Delta_0=\lambda \Delta_0
\frac{2}{\pi}\arctan\left(\frac{W}{2\pi T_c}\right).
\label{w}
\end{eqnarray}
Inverting Eq. (\ref{w}) we obtain thus, for $T_c \gg W$,
$T_c =(W/2\pi)/\tan(\pi/2\lambda)\approx \lambda W/\pi^2$,
showing that Eq. (\ref{my}) is valid as far as $\omega_0 \ll T_c \ll W$,
whereas for $T_c \gg W$ the divergence at $\lambda_c$ is removed and
a linear behavior as function of $\lambda$ is achieved.

\subsection{Superconducting Matsubara gap $\Delta_{\rm M}$}

Let us now investigate the asymptotic behavior
of the zero temperature Matsubara gap $\Delta_{\rm M}$.
Once more, we assume the limit $\Delta_{\rm M} \gg \omega_0$ 
and we shall check later the consistency of this ansatz.
To get an analytical expression for $\Delta_{\rm M}$,
we essentially follow Refs. \onlinecite{carbotte_rmp,cmm}.
Transforming, in the zero temperature limit, the Matsubara
sum in an integral, $T\sum_m \rightarrow \int d\omega/2\pi$,
we can write:
\begin{eqnarray}
Z(\omega)
&=&
1+\frac{\lambda\eta}{2\omega}\int
\frac{d\omega'\omega'}{\sqrt{\omega'^2+\Delta^2(\omega')}}
\frac{\omega_0^2}{\omega_0^2+(\omega-\omega')^2},
\label{zapp}
\end{eqnarray}
\begin{eqnarray}
Z(\omega)\Delta(\omega)
&=&
\frac{\lambda}{2}\int
\frac{d\omega' \Delta(\omega')}
{\sqrt{\omega'^2+\Delta^2(\omega')}}
\frac{\omega_0^2}{\omega_0^2+(\omega-\omega')^2}.
\label{gapapp}
\end{eqnarray}
We also employ the simple model:\cite{carbotte_rmp,cmm}
\begin{equation}
\Delta(\omega)=\left\{
\begin{array}{ll}
\Delta_{\rm M} & \hspace{5mm}\mbox{for $\omega < \alpha \Delta_{\rm M}$}, \\
0 & \hspace{5mm}\mbox{for $\omega > \alpha \Delta_{\rm M}$},
\end{array}
\right.
\end{equation}
which we have tested numerically to be appropriate even for
unbalanced superconductors.
Here $\alpha$ is a constant of the order of unity.
Using this model and expanding Eqs. (\ref{zapp})-(\ref{gapapp})
in powers of $\omega_0/\Delta_{\rm M}$ we end up with:
\begin{eqnarray}
1+\eta \lambda\frac{\pi \omega_0}{2 \Delta_{\rm M}}
-c_1\eta \lambda\frac{\omega_0^2}{\Delta_{\rm M}^2}
&=&
\lambda\frac{\pi \omega_0}{2 \Delta_{\rm M}}
-c_2\lambda\frac{\omega_0^2}{\Delta_{\rm M}^2},
\label{gapoo2}
\end{eqnarray}
where $c_1$, $c_2$ are constant factors depending on $\alpha$.
For $\alpha=1$ we have for instance $c_1=2(\sqrt{2}-1/2)$,
$c_2=\sqrt{2}$,\cite{cmm}
while for $\alpha=2$ we have $c_1=\sqrt{5}-1$,
$c_2=\sqrt{5}/2$.\cite{carbotte_rmp}

In the perfectly balanced case $\eta=1$ the linear terms
in $\omega_0/\Delta_{\rm M}$ in Eq. (\ref{gapoo2}) cancel out,
and one recovers the usual result $\Delta_{\rm M}^{\eta=1}=\sqrt{\lambda}
\sqrt{c_1-c_2}\omega_0$ (we remind that $c_1-c_2$ is a positive quantity).
On the other hand, for $\eta<1$ the solution of Eq. (\ref{gapoo2})
is dominated by the linear terms giving:
\begin{eqnarray}
\Delta_{\rm M}^{\eta<1}
&=&
\lambda(1-\eta)\frac{\pi \omega_0}{2}.
\end{eqnarray}
More complex is the case $\eta>1$, where retaining only the linear term
is not sufficient. In this case one needs to consider explicitly
also the $(\omega_0/\Delta_{\rm M})^2$ and to solve the quadratic
equation (\ref{gapoo2}).
We obtain:
\begin{eqnarray}
\left(\frac{\Delta_{\rm M}}{\omega_0}\right)^2
+\frac{\lambda\pi(\eta-1)}{2}\left(\frac{\Delta_{\rm M}}{\omega_0}\right)
-\lambda(\eta c_1-c_2)=0,
\end{eqnarray}
so that
\begin{eqnarray}
\frac{\Delta_{\rm M, max}^{\eta>1}}{\omega_0}
&=&
\frac{1}{2}
\left[
-\frac{\lambda\pi(\eta-1)}{2}
\right.
\nonumber\\
&&\left.
+\sqrt{
\frac{\lambda^2\pi^2(\eta-1)^2}{4}
+4\lambda(\eta c_1-c_2)
}
\right]
\nonumber\\
&\simeq&
\frac{4(\eta c_1-c_2)}{2\pi(\eta-1)},
\label{gapetabig}
\end{eqnarray}
showing that for $\eta>1$ $\Delta_{\rm M}$ saturates at a finite value
for $\lambda\rightarrow \infty$, and Eq. (\ref{gapetabig})
can be considered an
upper limit for it.

\subsection{Temperature dependence of $\Delta_{\rm M}(T)$}

In this section we investigate the temperature
behavior of the Matsubara gap $\Delta_{\rm M}(T)$ close to $T_c$.
We assume once more th limit $T_c \gg \omega_0$ and thus the
validity of the one-Matsubara-gap model.
Within these approximation we can rewrite Eqs. (\ref{Z})
as:
\begin{eqnarray}
Z_0
&=&
1+ \eta\lambda
\sum_{m=0,-1} 
\frac{\omega_0^2}{\omega_0^2+(\omega_n-\omega_m)^2}
\frac{\omega_m}{\sqrt{\omega_m^2+\Delta_m^2}}
\nonumber\\
&&
+\sum_{m\neq 0,-1}
\frac{\omega_0^2}{\omega_0^2+(\omega_n-\omega_m)^2}
\mbox{sgn}[\omega_m]\nonumber\\
&=&
1+\frac{\eta\lambda\pi T}{\sqrt{\pi^2 T^2+\Delta_0^2}}
\left(1-\frac{\omega_0^2}{4\pi^2T^2}\right)
+\frac{\eta\lambda\omega_0^2}{4\pi^2T^2},
\label{app-z}
\end{eqnarray}
where we have used $\Delta_0=\Delta_{-1}$,
and in similar way,
\begin{eqnarray}
Z_0\Delta_0
&=&
\lambda \pi T 
\frac{\Delta_0}{\sqrt{\pi^2 T^2+\Delta_0^2}}
\left(1+\frac{\omega_0^2}{4\pi^2T^2}\right).
\label{app-d}
\end{eqnarray}
Plugging (\ref{app-z}) in (\ref{app-d}) we have:
\begin{eqnarray}
\Delta_0
&=&
\lambda(1-\eta)
\frac{\pi T\Delta_0}{\sqrt{\pi^2 T^2+\Delta_0^2}}
\nonumber\\
&&
+\lambda(1+\eta)
\frac{\pi T\Delta_0}{\sqrt{\pi^2 T^2+\Delta_0^2}}
\frac{\omega_0^2}{4\pi^2T^2}
-\frac{\eta\lambda\omega_0^2\Delta_0}{4\pi^2T^2},
\end{eqnarray}
which we can rewrite as:
\begin{eqnarray}
\frac{\sqrt{\pi^2 T^2+\Delta_0^2}}{\pi T}
&=&
\frac{
\lambda(1-\eta) +\lambda(1+\eta)
\frac{\displaystyle \omega_0^2}{\displaystyle 4\pi^2 T^2}}
{\left(1+\frac{\displaystyle \eta\lambda\omega_0^2}{\displaystyle 4\pi^2T^2}\right)}.
\label{gapexp}
\end{eqnarray}

Expanding left- and right-hand sides of Eq. (\ref{gapexp}) at the
second order in $\Delta$ and at the linear order of $\delta=1-T/T_c$,
we have:
\begin{eqnarray}
\frac{\Delta_0^2}{2\pi^2T_c^2}
&=&
\frac{2\delta \lambda \omega_0^2}{4\pi^2 T_c^2+\eta\lambda\omega_0^2}
\nonumber\\
&=&
2\delta
\frac{1-\lambda(1-\eta)}{2-\lambda(1-\eta)},
\label{app-gap-t}
\end{eqnarray}
where we made use of Eq. (\ref{app-tc}).
Eq. (\ref{app-gap-t}) reduces to the standard relation
$(\Delta_0/\pi T_c)^2=2\delta$ in the perfectly balanced
case $\eta=1$.\cite{carbotte_rmp}
On the other hand, for $\eta<1$ we have
\begin{eqnarray}
\frac{\Delta_0^2}{\pi^2T_c^2}
&=&
4\delta
\frac{\lambda_c-\lambda}{2\lambda_c-\lambda},
\end{eqnarray}
which shows that $\Delta(T) \simeq c \sqrt{T_c-T}$ (for $T \simeq T_c$)
with a
vanishing coefficient $c$ for $\lambda\rightarrow \lambda_c$.
Finally, for $\eta > 1$, Eq. (\ref{app-gap-t}) is well-behaved
in the limit $\lambda\rightarrow \infty$ and it gives
\begin{eqnarray}
\frac{\Delta_0^2}{\pi^2T_c^2}
&=&
4\delta,
\end{eqnarray}
with a coefficient $c$ twice larger than the usual.

Eq. (\ref{gapexp}) can be employed also to study the temperature behavior
of the superconducting gap in the regime $\eta<1$ and $\lambda > \lambda_c$,
where the system is superconducting even at high temperature and
no finite $T_c$ is predicted. Performing the limit $T\gg \omega_0$ we obtain:
\begin{eqnarray}
\frac{\sqrt{\pi^2 T^2+\Delta_0^2}}{\pi T}
&=&
\lambda(1-\eta),
\end{eqnarray}
and
\begin{eqnarray}
\Delta_0 \simeq \pi T \sqrt{(\lambda/\lambda_c)^2-1},
\end{eqnarray}
showing that $\Delta_0$ increases linearly with $T$
in this regime.


\begin{thebibliography}{99}
\bibitem{note-hc}
A similar change of curvature in $H_c(T)$
was observed also in Ref. \onlinecite{mwc} in the
strong-coupling asymptotic limit $\lambda\rightarrow \infty$ 
of perfectly balanced $\eta=1$ superconductors.
\end{thebibliography}
\end{document}